\def \VersionLong {}
\def \VersionAuthor {}
\ifdefined\VersionAuthor
	\newcommand{\AuthorVersion}[1]{#1}
	\newcommand{\FinalVersion}[1]{}
\else
	\newcommand{\AuthorVersion}[1]{}
	\newcommand{\FinalVersion}[1]{#1}
\fi

\documentclass[a4paper,10pt]{llncs}
\makeatletter
\AtBeginDocument{%
  \@ifpackageloaded{hyperref}
  {\def\@doi#1{\href{https://doi.org/#1}
      {\ttfamily https://doi.org/#1}\egroup}}
  {\def\@doi#1{\ttfamily https://doi.org/#1\egroup}}
  \def\doi{\bgroup\catcode`\_=12\relax\@doi}}
\makeatother

\usepackage[utf8]{inputenc}
\usepackage[english]{babel}

\usepackage[ruled,vlined,linesnumbered]{algorithm2e}
	\SetKwInOut{Input}{input}
	\SetKwInOut{Output}{output}

\usepackage{subcaption}

\usepackage{paralist} %

\newenvironment{ienumerate}
	{\ifdefined\VersionLong\begin{enumerate}\else\begin{inparaenum}[\itshape i\upshape)]\fi}
	{\ifdefined\VersionLong\end{enumerate}\else\end{inparaenum}\fi}

\newenvironment{oneenumerate}
	{\ifdefined\VersionLong\begin{enumerate}\else\begin{inparaenum}[1)]\fi}
	{\ifdefined\VersionLong\end{enumerate}\else\end{inparaenum}\fi}

\usepackage{amsmath} %
\usepackage{amssymb} %

\usepackage[misc,geometry]{ifsym} %

\usepackage{csquotes}

\ifdefined \VersionLong
	\newcommand{\LongVersion}[1]{#1}
	\newcommand{\ShortVersion}[1]{}
\else
	\newcommand{\LongVersion}[1]{}
	\newcommand{\ShortVersion}[1]{#1}
\fi

\ifdefined\VersionAuthor
	\usepackage[backend=biber,backref=true,style=alphabetic,url=true,doi=true,defernumbers=true,sorting=anyt,maxnames=99]{biblatex} %
	\renewbibmacro*{doi+eprint+url}{%
		\iftoggle{bbx:doi}
			{\color{black!40}\footnotesize\printfield{doi}}
			{}%
		\newunit\newblock
		\iftoggle{bbx:eprint}
			{\usebibmacro{eprint}}
			{}%
		\newunit\newblock
		\iftoggle{bbx:url}
			{\usebibmacro{url+urldate}}
			{}%
	}

\fi
\ifdefined\VersionWithComments
	\usepackage{draftwatermark}
	\SetWatermarkText{draft}
	\SetWatermarkScale{15}
	\SetWatermarkColor[gray]{0.9}
\fi
\usepackage[svgnames,table]{xcolor}
\definecolor{darkblue}{rgb}{0, 0, 0.7}

\usepackage[
		pdfauthor={André, Marinho, van de Pol},%
		pdftitle={A benchmarks library for extended parametric timed automata},
		breaklinks  = true,
		colorlinks  = true,
	\ifdefined \VersionWithComments
	\fi
		citecolor   = blue!50!blue,
		linkcolor   = darkblue,
		urlcolor    = blue!50!green,
	]{hyperref}

\usepackage[capitalise,english,nameinlink]{cleveref} %
\crefname{line}{\text{line}}{\text{lines}} %

\usepackage{tikz}
\usetikzlibrary{arrows,automata}
\tikzstyle{pta}=[auto, ->, >=stealth']
\tikzstyle{every node}=[initial text=]
\tikzstyle{location}=[rectangle, rounded corners, minimum size=12pt, draw=black, fill=blue!10, inner sep=2pt]
\tikzstyle{invariant}=[draw=black, dotted, inner sep=1pt] %
\tikzstyle{final}=[double, fill=blue!50]

\tikzstyle{urgent}=[fill=yellow, thick, dotted] %

\definecolor{coloract}{rgb}{0.50, 0.70, 0.30}
\definecolor{colorclock}{rgb}{0.4, 0.4, 1}
\definecolor{colordisc}{rgb}{1, 0, 1}
\definecolor{colorloc}{rgb}{0.4, 0.4, 0.65}
\definecolor{colorparam}{rgb}{1, 0.6, 0.0}

\definecolor{loccolor1}{rgb}{1, 0.3, 0.3}
\definecolor{loccolor2}{rgb}{0.3, 1, 0.3}
\definecolor{loccolor3}{rgb}{0.3, 0.3, 1}
\definecolor{loccolor4}{rgb}{1, 0.3, 1}
\definecolor{loccolor5}{rgb}{1, 1, 0.3}
\definecolor{loccolor6}{rgb}{0.3, 1, 1}
\definecolor{loccolor7}{rgb}{0.9, 0.6, 0.2}
\definecolor{loccolor8}{rgb}{0.7, 0.4, 1}
\definecolor{loccolor9}{rgb}{0.5, 1, 0.75}
\definecolor{loccolor10}{rgb}{0.8, 0.7, 0.6}
\definecolor{loccolor11}{rgb}{0.6, 0.7, 0.8}
\definecolor{loccolor12}{rgb}{0.2, 0.5, 0.9}
\definecolor{loccolor13}{rgb}{0.5, 0.9, 0.2}
\definecolor{loccolor14}{rgb}{0.9, 0.2, 0.5}
\definecolor{loccolor15}{rgb}{0.7, 0.7, 0.7}
\definecolor{loccolor16}{rgb}{0.8, 0.8, 0.5}

\newcommand{\styleact}[1]{\ensuremath{\textcolor{coloract}{{#1}}}}
\newcommand{\styleclock}[1]{\ensuremath{\textcolor{colorclock}{{#1}}}}
\newcommand{\styledisc}[1]{\ensuremath{\textcolor{colordisc}{\mathrm{#1}}}}
\newcommand{\styleloc}[1]{\ensuremath{\mathsf{#1}}}
\newcommand{\styleparam}[1]{\ensuremath{\textcolor{colorparam}{{#1}}}}

\newcommand{\styleBen}[1]{\texttt{#1}}

\newcommand{\rowHeader}{\rowcolor{blue!20}}

\newcommand{\cellYes}{\cellcolor{green!40}\textbf{$\surd$}}
\newcommand{\cellNo}{\cellcolor{red!40}\textbf{$\times$}}

\ifdefined\VersionWithComments
	\usepackage[colorinlistoftodos,textsize=footnotesize]{todonotes}
\else
	\usepackage[disable]{todonotes}
\fi
\newcommand{\gennote}[3]{\todo[size=\scriptsize,linecolor=#2,backgroundcolor=#2!25,bordercolor=#2]{#3: #1}\xspace}
\newcommand{\ea}[1]{\gennote{#1}{blue}{ÉA}}
\newcommand{\dm}[1]{{\gennote{#1}{purple}{DM}}}

\ifdefined \VersionWithComments
	\newcommand{\todoinline}[1]{\mbox{}{\color{red}{\textbf{TODO}\ifx#1\\\else:\ \fi #1}}} %
\else
	\newcommand{\todoinline}[1]{}
\fi

\usepackage{verbatim} %

\newcommand{\init}{_0}

\newcommand{\A}{\ensuremath{\mathcal{A}}}

\newcommand{\assign}{\leftarrow}

\newcommand{\loc}{\ensuremath{\ell}} %
\newcommand{\locinit}{\loc\init}
\newcommand{\locfinal}{\ensuremath{\loc_f}}

\newcommand{\param}{p} %
\newcommand{\R}{{\mathbb{R}}}

\newcommand{\setN}{\ensuremath{\mathbb N}}

\newcommand{\setR}{\ensuremath{\mathbb R}}
\newcommand{\setRplus}{\ensuremath{\setR_{+}}} %

\newcommand{\actDone}{\styleact{\mathit{done}}}
\newcommand{\actDrink}{\styleact{\mathit{drink}}}
\newcommand{\actRestart}{\styleact{\mathit{restart}}}
\newcommand{\clockx}{\ensuremath{\styleclock{x}}} %
\newcommand{\clocky}{\ensuremath{\styleclock{y}}} %
\newcommand{\clockt}{\ensuremath{\styleclock{t}}}
\newcommand{\MAXCOFFEE}{\styleparam{\mathit{max}}}
\newcommand{\nb}{\styledisc{\mathit{nb}}}
\newcommand{\paramCoffee}{\styleparam{p_\mathit{coffee}}}
\newcommand{\paramNeed}{\styleparam{p_\mathit{need}}}
\newcommand{\paramTotal}{\styleparam{p_\mathit{total}}}

\newcommand{\hytech}{\textsc{HyTech}}
\newcommand{\imitator}{\textsf{IMITATOR}}

\newcommand{\uppaal}{\textsc{Uppaal}}

\newcommand{\benchmark}[1]{\texttt{#1}}

\ifdefined \VersionWithComments
 	\definecolor{colorok}{RGB}{80,80,150}
\else
	\definecolor{colorok}{RGB}{0,0,0}
\fi

\newcommand{\eg}{\textcolor{colorok}{e.g.,}\xspace}

\newcommand{\ie}{\textcolor{colorok}{i.e.,}\xspace}

\makeatletter
\def\orcidID#1{\smash{\href{https://orcid.org/#1}{\protect\raisebox{-1.25pt}{\protect\includegraphics{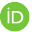}}}}}
\makeatother

\usepackage[firstpage]{draftwatermark}
\SetWatermarkText{\hspace*{100mm}\raisebox{180mm}{\includegraphics{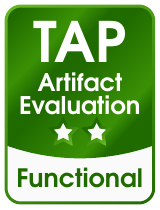}}}
\SetWatermarkAngle{0}

\title{A Benchmarks Library for Extended Parametric Timed Automata\todo{This is the version with comments. To disable comments, comment out line~3 in the \LaTeX{} source.}\thanks{%
	\AuthorVersion{
	This is the author (and extended) version of the manuscript of the same name published in the proceedings of the  15th International Conference on Tests and Proofs (\href{https://www.univ-orleans.fr/lifo/evenements/TAP2021/}{TAP 2021}).
	The final authenticated version is available at
		\href{https://www.doi.org/10.1007/978-3-030-79379-1_3}{\nolinkurl{10.1007/978-3-030-79379-1_3}}.
	}%
	This work is partially supported by the ANR-NRF French-Singaporean research program \href{https://www.loria.science/ProMiS/}{ProMiS} (ANR-19-CE25-0015).
}
}
\author{\'Etienne Andr\'e\inst{1}%
	\orcidID{0000-0001-8473-9555}
\and
	Dylan Marinho\inst{1}%
	\orcidID{0000-0002-2548-6196}%
	${}^{\href{https://github.com/DylanMarinho}{\text{\Letter}}}$
\and
	\\
	Jaco van de Pol\inst{2}%
	\orcidID{0000-0003-4305-0625}
}
\institute{%
	Université de Lorraine, CNRS, Inria, LORIA, F-54000 Nancy, France
\and
	Department of Computer Science, Aarhus University, Aarhus, Denmark
}

\begin{document}
\sloppy

\AuthorVersion{
	\pagestyle{plain}
}

\maketitle

\setcounter{footnote}{0}

\begin{abstract}
	Parametric timed automata are a powerful formalism for reasoning on concurrent real-time systems with unknown or uncertain timing constants.
	In order to test the efficiency of new algorithms, a fair set of benchmarks is required.
	We present an extension of the \imitator{} benchmarks library, that accumulated over the years a number of case studies from academic and industrial contexts.
	We extend here the library with several dozens of new benchmarks; these benchmarks highlight several new features: liveness properties, extensions of (parametric) timed automata (including stopwatches or multi-rate clocks), and unsolvable toy benchmarks.
	These latter additions help to emphasize the limits of state-of-the-art parameter synthesis techniques, with the hope to develop new dedicated algorithms in the future.

	\LongVersion{}
\end{abstract}
\section{Introduction}\label{section:introduction}

Timed automata (TAs)~\cite{AD94} are a powerful formalism for reasoning on concurrent real-time systems.
Their parametric extension (\emph{parametric timed automata}, PTAs~\cite{AHV93}) offer the use of \emph{timing parameters} (unknown or uncertain timing constants), allowing to verify properties on a model at an earlier design stage, or when the exact values of constants at runtime may be unknown.
The model checking problem with its binary answer (``yes''/``no'') becomes the \emph{parameter synthesis} problem: ``for which values of the parameters does the model satisfy its specification?''.

In the past few years, a growing number of new synthesis algorithms were proposed for PTAs, \eg{} using
	bounded model-checking~\cite{KP12},
	compositional verification~\cite{ABBCR16,ALin17},
	distributed verification~\cite{ACN15},
	for liveness properties~\cite{BBBC16,NPP18,AAPP21},
	for dedicated problems~\cite{CPR08}---notably for testing timed systems~\cite{FK13,Andre16,LSBL17,AAGR19,LGSBL19,AAPP21}.
However, these works consider different benchmarks sets, making it difficult to evaluate which technique is the most efficient for each application domain.

A benchmarks suite for (extended) PTAs can be used for different purposes:
\begin{ienumerate}
	\item when developing new algorithms for (extensions of) PTAs and testing their efficiency by comparing them with existing techniques;
	\item when evaluating benchmarks for extensions of TAs~\cite{AD94} (note that valuating our benchmarks with a parameter valuation yields a timed or multi-rate automaton);
	and
	\item when looking for benchmarks fitting in the larger class of \emph{hybrid} automata~\cite{ACHH92}.
\end{ienumerate}

\paragraph{Contribution}
In~\cite{Andre18FTSCS,WebIMIlib1}, we introduced a first library of 34 benchmarks\LongVersion{, 80 models} and 122 properties, for PTAs.
However, this former \LongVersion{version of the }library suffers from several issues.
First, its syntax is only compatible with the syntax of version \href{https://github.com/imitator-model-checker/imitator/releases/tag/v2.12}{2.12} of \imitator{}~\cite{AFKS12}, while \imitator{} recently shifted to
version \href{https://github.com/imitator-model-checker/imitator/releases/tag/v3.0.0}{3.0}~\cite{Andre21}, with a different calling paradigm.\footnote{%
	While many keywords remain the same in the model, the property syntax has been completely rewritten, and the model checker now takes as input a model file \emph{and} a property file.
	In addition, new properties are now possible, and the syntax has been extended with some useful features such as multi-rate clocks\LongVersion{ or ``if-then-else'' control structures}.
}
Second, the former version contains exclusively safety/reachability properties (plus some ``robustness'' computations\LongVersion{, using the trace preservation synthesis algorithm}~\cite{ALM20}).
Third, only syntactic information is provided (benchmarks, metrics on the benchmarks), and no semantic information (expected result, approximate computation time, and \LongVersion{approximate }number of states to explore).

In this work, we extend our former library with a list of new features, including syntactic extensions (notably multi-rate clocks~\cite{ACHHHNOSY95});
we also focus on \emph{unsolvable} case studies, \ie{} simple examples for which no known algorithm allows computation of the result, with the ultimate goal to encourage the community to address these cases.
In addition, we add \emph{liveness} properties\LongVersion{, \ie{} cycle synthesis}.
Also, we add \emph{semantic} criteria, with an approximate computation time for the properties, an expected result (whenever available) and an approximate number of explored \LongVersion{symbolic }states\LongVersion{ (computed using \imitator{})}.
The rationale is to help users by giving them an idea of what to expect for each case study.
Also, our consolidated classification aims at helping tool developers to select within our library which benchmarks suit them (\eg{} ``PTAs without stopwatches, with many locations and a large state space'').

To summarize, we propose a new version of our library enhancing the former one as follows:
\begin{enumerate}
	\item adding 22 new benchmarks (39 models)
	\begin{itemize}
		\item adding benchmarks for \emph{liveness} properties;
		\item adding a set of toy \emph{unsolvable} benchmarks, to emphasize the limits of state-of-the-art parametric verification techniques, and to encourage the community to develop new dedicated algorithms in the future;
	\end{itemize}
	\item refactoring all existing benchmarks, so that they now implement the syntax of the 3.0 version of \imitator{};
	\item providing a better classification of benchmarks;
	\item highlighting extensions of (parametric) timed automata, such as multi-rate clocks~\cite{ACHHHNOSY95}, stopwatches~\cite{CL00}, …
	\item offering an automated translation of our benchmarks to the new JANI~\cite{BDHHJT17,WebJANI}
	model interchange format, offering a unified format for quantitative automata-based formalisms.
	This way, the library can be used by any tool using JANI as an input format, and supporting (extensions of) TAs.
	Even though other tools implementing the JANI formalism do not handle parameters, they can run on \emph{instances} of our benchmarks, \ie{} by valuating the PTAs with concrete valuations of the parameters.
\end{enumerate}

\begin{table}[tb]
	\caption{Selected new features}\label{table:comparison}
	\scriptsize
	\centering
	\setlength{\tabcolsep}{1.5pt} %
	\begin{tabular}{| r | r | r | r | c | c | l | c | c | c |c |c |c |}
		\hline
		\rowHeader{}Library & \multicolumn{3}{c|}{Size} & \multicolumn{2}{c|}{Metrics} & \multicolumn{2}{c|}{Format} & Categories & \multicolumn{3}{c|}{Properties} & Analysis
		\\
		\hline
		\rowHeader{}Version & Bench\LongVersion{marks}\ShortVersion{.} & Models & Prop\LongVersion{erties}\ShortVersion{.} & Static & Semantic & \texttt{.imi} & JANI & Unsolvable & EF & TPS & liveness & Results
		\\
		\hline
		1.0 \cite{WebIMIlib1} & 34 & 80 & 122 & \cellYes{} & \cellNo{} & 2.12 & \cellNo{} & \cellNo{} & \cellYes{} & \cellYes{} & \cellNo{} & \cellNo{}
		\\
		\hline
		2.0 \cite{WebIMIlib2} & 56 & 119 & 216 & \cellYes{} & \cellYes{} & 3.0 & \cellYes{} & \cellYes{} & \cellYes{} & \cellYes{} & \cellYes{} & \cellYes{}
		\\
		\hline
	\end{tabular}

\end{table}

We summarize the most significant dimensions of our extension in \cref{table:comparison}.
EF (using the TCTL syntax) denotes reachability/safety, and TPS (``trace preservation synthesis'') denotes robustness analysis.

\paragraph{Outline}
We discuss related libraries in \cref{section:related}.
We briefly recall \imitator{} PTAs in \cref{section:pta}.
We present our library in \cref{section:library}, and we give perspectives in \cref{section:perspectives}.

\section{Related Libraries}\label{section:related}

RTLib~\cite{SGQ16} is a library of real-time systems modeled as timed automata.
Contrary to our \LongVersion{goal and proposed }solution, it does not consider parametric models.

Two hybrid systems benchmarks libraries were proposed in~\cite{FI04,CSBMAFK15}.
Despite being more expressive than PTAs in theory, these formalisms cannot be compared in practice: most of them do not refer to timing parameters.
Moreover, these libraries only focus on reachability properties\LongVersion{ and non-parameterized benchmarks}.

The PRISM benchmark suite~\cite{KNP12} collects probabilistic models and properties.
Despite including some timing aspects, time is not the focus there.
The collection of Matlab/Simulink models~\cite{HAF14} focuses on timed model checking, but has
no parametric extensions.
Two of our benchmarks (\benchmark{accel} and \benchmark{gear}) originate from a translation of their models to (extensions of) PTAs~\cite{AHW18}.

The JANI specification \cite{BDHHJT17} defines a representation of automata with quantitative extensions and variables\LongVersion{; their syntax is supported by several verification tools}.
A library of JANI benchmarks is also provided; such benchmarks come from PRISM, Modest, Storm and FIG, and therefore cannot be applied to parameter synthesis for timed systems.

Also, a number of model checking competitions started in the last two decades, accumulating over the years a number of sets of benchmarks, such as
	the ARCH\LongVersion{ (Applied Verification for Continuous and Hybrid Systems)} ``friendly competition''~\cite{FAABBCGGMM19,WebARCH},
	the Petri Nets model checking contest\LongVersion{ (MCC)}~\cite{ABCDGHHJ19,WebMCC},
	the MARS\LongVersion{ (Models for Formal Analysis of Real Systems)} workshop repository~\cite{WebMARS},
	the WATERS workshop series~\cite{WebWATERS15},
	etc.

\LongVersion{
	In~\cite{Andre18FTSCS} we introduced a first version of our parametric timed automata library~\cite{WebIMIlib1}, including academic benchmarks, industrial case studies and toy case studies.
}

\LongVersion{\paragraph{Position of our Library}}
Our library aims at providing benchmarks for \emph{parameter} synthesis for (extensions of) TAs.
Notably, we go beyond the TA syntax (offering some benchmarks with multi-rate clocks, stopwatches, timing parameters, additional global variables), while not offering the full power of hybrid automata (differential equations, complex flows).
To the best of our knowledge, no other set of benchmarks addresses specifically the \emph{synthesis} of timing parameters.

\section{Parametric Timed Automata}\label{section:pta}

\paragraph{Parametric Timed Automata (PTAs)}
Timed automata (TAs)~\cite{AD94} extend finite-state automata with \emph{clocks}, \ie{} real-valued variables evolving at the same rate~1, that can be compared to integers along edges (``guards'') or within locations (``invariants'').
Clocks can be reset (to~0) along transitions.
PTAs extend TAs with (timing) parameters, \ie{} unknown rational-valued constants~\cite{AHV93}.
These timing parameters can have two main purposes:
\begin{itemize}
	\item model unknown constants, and \emph{synthesize} suitable values for them, at an early design stage; or
	\item verify the system for a full range of constants, as their actual value may not be exactly known before execution; this is notably the case of the FMTV Challenge by Thales\LongVersion{\footnote{%
		\url{https://www.imitator.fr/data/FMTV15/FMTV-2015-Challenge.pdf}
	}} at WATERS~2015~\cite{WebWATERS15}
	that features periods known with a limited precision only (\ie{} constant but of unknown exact value), and that we were able to solve using parametric timed automata~\cite{SAL15}.
		(This benchmark \benchmark{FMTV1} is part of our library.)
\end{itemize}
PTAs can be synchronized together on shared actions, or by reading shared variables.
That is, it is possible to perform the parallel composition of several PTAs, using a common actions alphabet.
This allows users to define models component by component.

\begin{example}\label{example:PTAs}
	Consider the toy PTA in \cref{fig:unsolvable-N}.
	It features two clocks $\clockx$ and~$\clocky$, one parameter $\styleparam{\param}$ and two locations.
	$\loc_0$ is the initial location, while $\locfinal$ is accepting.
	The \emph{invariant}, defining the condition to be fulfilled to remain in a location, is depicted as a dotted box below the location (\eg{} $\clockx \leq 1$ in~$\loc_0$).
	A transition from~$\loc_0$ to~$\locfinal$ can be taken when its \emph{guard} (``$\clockx = 0 \land \clocky = \styleparam{\param}$'') is satisfied; the other transition (looping on~$\loc_0$) can be taken whenever $\clockx = 1$, resetting $\clockx$ to~0.

	Observe that, if $\styleparam{\param} = 0$, then the guard $\clockx = 0 \land \clocky = 0$ is immediately true, and~$\locfinal$ is reachable in 0-time.
	If $\styleparam{\param} = 1$, the guard becomes $\clockx = 0 \land \clocky = 1$, which is not initially satisfied, and one needs to loop once over~$\loc_0$, yielding $\clockx = 0 \land \clocky = 1$, after which $\locfinal$ is reachable.
	In fact, it can be shown that the answer to the reachability synthesis problem ``synthesize the valuations of~$\styleparam{\param}$ such that $\locfinal$ is reachable'' is exactly $\styleparam{\param} = i, i \in \setN$.
\end{example}

\paragraph{Extending the PTAs Syntax}
Our library follows the \imitator{} syntax.
Therefore, some benchmarks (clearly marked as such) go beyond the traditional PTAs syntax, and are referred to IPTAs (\imitator{} PTAs).
These extensions include:
\begin{description}
	\item[Urgent locations]
	Locations where time cannot elapse.

	\item[Global rational-valued variables]
	Such ``discrete'' variables can be updated along transitions, and can also be part of the clock guards and invariants.
	\LongVersion{This is the case of \styledisc{nb} in \cref{figure:example-researcher}.}

	\item[Arbitrary flows]
	Some benchmarks require arbitrary (constant) flows for clocks; this way, clocks do not necessary evolve at the same time, and can encode different concepts from only time, \eg{} temperature, amount of completion, continuous cost.
		Their value can increase or decrease at any predefined rate in each location, and can become negative.
		In that sense, these clocks are closer to \emph{continuous variables} (as in hybrid automata) rather than TAs' clocks; nevertheless, they still have a constant flow, while hybrid automata can have more general flows.
		This makes some of our benchmarks fit into a parametric extension of \emph{multi-rate automata}~\cite{ACHHHNOSY95}.
		This notably includes stopwatches, where clocks can have a 1 or 0-rate~\cite{CL00}.
		\LongVersion{In \cref{figure:example-researcher}, \clockx{} has rate~2 in \styleloc{workingFast}, and is stopped in \styleloc{coffee} (rate~0, or stopwatch).}
\end{description}

\LongVersion{
\begin{example}\label{example:researcher}
	An example of IPTA following some of the extensions offered by our library is given in\LongVersion{~\cref{figure:example-researcher}.
	A full description of this example is given in} \cref{example:researcher:full} in \cref{appendix:example}.
\end{example}
}

\section{The Benchmarks Library}\label{section:library}
\subsection{Organization}\label{subsection:organization}

The library is made of a set of \emph{benchmarks}.
Each benchmark may have different \emph{models}: for example, \styleBen{Gear} comes with ten models, of different sizes (the number of locations notably varies), named \styleBen{Gear:1000} to \styleBen{Gear:10000}.
Similarly, some \styleBen{Fischer} benchmarks come with several models, each of them corresponding to a different number of processes.
Finally, each model comes with one or more \emph{properties}.
For example, for \styleBen{Gear:2000}, one can run either reachability synthesis, or minimal reachability synthesis.

The benchmark library, in its 2.0 version, covers 56 benchmarks, which group 119 models and 216 properties\dm{The "number of properties" here stands for the number of (model,prop), not the number of *.imiprop files}.

From the previous version \cite{Andre18FTSCS}, 39 models have been added: %
beyond all \benchmark{Unsolvable} models, and a few more additions, we notably added a second model of the Bounded Retransmission Protocol (\benchmark{BRPAAPP21}), recently proposed in~\cite{AAPP21}. %

Benchmarks come from industrial collaborations (\eg{} with Thales, ST-Microelectronics, ArianeGroup, Astrium), from academic papers from different communities (\eg{} real-time systems, monitoring, testing)\ describing case studies, and from our experience in the field (notably the ``unsolvable'' benchmarks).
For benchmarks extracted from published works, a complete bibliographic reference is given.

\subsection{Distribution}\label{subsection:distribution}
\begin{figure}[tb]
	\centering
	\includegraphics[width=\linewidth]{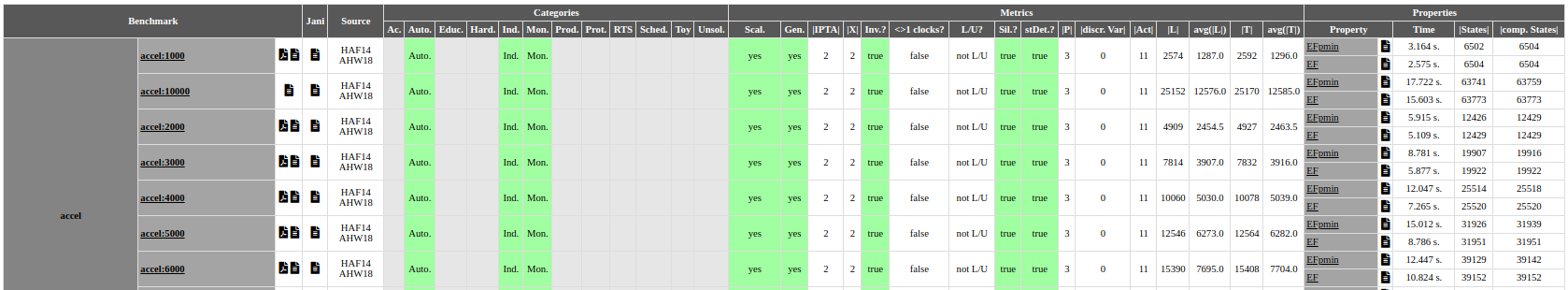}
	\caption{The \imitator{} benchmark library Web page}
	\label{fig:webpage-screenshot}
\end{figure}

The benchmark library is presented on a Web page available at \cite{WebIMIlib2} and permanently available at~\cite{AMP21data}.
Several columns (metrics, syntax used, categories, properties) allow users to select easily which benchmarks fit their need (see \cref{fig:webpage-screenshot}).

Our benchmarks are distributed in the well-documented \imitator{} 3.0.0 input format~\cite{imitatorUserManual}, which is a \emph{de facto} standard for PTAs.
\imitator{} can provide automated translations to the non-parametric timed model checker \uppaal{}~\cite{LPY97}, as well as the hybrid systems model checker \hytech{}~\cite{HHW95} (not maintained anymore).
However, some differences (presence of timing parameters or complex guards in \imitator{}, difference in the semantics of the synchronization model) may not preserve the semantic equivalence of the models.

In addition, we offer all benchmarks in the JANI format~\cite{BDHHJT17}.
We recently implemented to this end (within \imitator{}) an automatic translation of IPTAs to their JANI specification.
Thus, all of our benchmarks can be fed to other verification tools supporting JANI as input.

All our benchmarks are released under the \href{https://creativecommons.org/licenses/by/4.0/}{CC by 4.0} license.

\subsection{Benchmarks Classification}\label{subsection:classification}

 For each benchmark, we provide multiple criteria, notably the following ones.
 \begin{description}
 	\item[Scalability] whether the models can be scaled according to some metrics, \eg{} the \benchmark{FischerPS08} benchmark can be scaled according to the number of processes competing for the critical section;
 	\item[Generation method] whether the models are automatically generated or not (\eg{} by a script, notably for scheduling real-time systems using PTAs, or to generate random words in benchmarks from the testing or monitoring communities); %
 	\item[Categorization] Benchmarks are tagged with one or more categories:
 	\begin{oneenumerate}
 		\item Academic,
 		\item Automotive,
 		\item Education,
 		\item Hardware,
 		\item Industrial,
 		\item Monitoring,
 		\item Producer-consumer,
 		\item Protocol,
 		\item Real-time system,
 		\item Scheduling,
 		\item Toy,
 		\item Unsolvable.
 	\end{oneenumerate} The proportion of each of these tags are given in \cref{table:categories} (the sum exceeds 100\,\% since benchmarks can belong to multiple categories).

\begin{table}[tb]
	\footnotesize
	\caption{Proportion of each category over the models}\label{table:categories}
	\centering

	\begin{tabular}{|l|r|r|}
		\hline
		\rowHeader{}Category & Number of models & Proportion \\ \hline
		All                  &              119 &    100\,\% \\ \hline
		Academic             &               54 &     45\,\% \\
		Automotive           &               20 &     17\,\% \\
		Education            &                9 &      8\,\% \\
		Hardware             &                6 &      5\,\% \\
		Industrial           &               33 &     28\,\% \\
		Monitoring           &               25 &     21\,\% \\
		ProdCons             &                5 &      4\,\% \\
		Protocol             &               34 &     29\,\% \\
		RTS                  &               46 &     39\,\% \\
		Scheduling           &                3 &      3\,\% \\
		Toy                  &               34 &     29\,\% \\
		Unsolvable           &               18 &     15\,\% \\ \hline
	\end{tabular}
\end{table}
 \end{description}

Moreover, we use the following static metrics to categorize our benchmarks:
 \begin{oneenumerate}
 	\item the numbers of PTA components (subject to parallel composition), of clocks, parameters, discrete variables and actions;
 	\item whether the benchmark has invariants, whether some clocks have a rate not equal to 1 (multi-rate/stopwatch) and silent actions (``$\epsilon$-transitions'');
 	\item whether the benchmark is an L/U-PTA\footnote{A subclass of PTAs where the set of parameters is partitioned into ``lower-bound'' and ``upper-bound'' parameters~\cite{HRSV02}.
 	L/U-PTAs enjoy nicer decidability properties.
 	}%
 	\LongVersion{ and strongly deterministic};
	\item the numbers of locations and transitions, and the total number of transitions.
 \end{oneenumerate}

\begin{table}[tb]
	\footnotesize
	\caption{Statistics on the benchmarks}\label{table:statistics}
	\noindent\begin{subtable}[b]{0.55\textwidth}

	\noindent\begin{tabular}{| l| r| r|}
		\hline
		\rowHeader{}Metric                & Average & Median                              \\ \hline
		Number of IPTAs        & 3       & 3                                   \\
		Number of clocks                  & 4       & 3                                   \\
		Number of parameters              & 4       & 3                                   \\
		Number of discrete variables      & 4       & 2                                   \\
		Number of actions                 & 12      & 11                                  \\
		Total number of locations         & 2004    & 22                                  \\
		\LongVersion{Locations per IPTA   & 979     & 5 \\}
		Total number of transitions & 2280 & 54 \\
		\LongVersion{Transitions per IPTA & 1067    & 13 \\}
	\hline                      
	\end{tabular}
	\end{subtable}
	~
	\begin{subtable}[b]{0.37\textwidth}
		\begin{tabular}{| l| r |}
			\hline
			\rowHeader{}Metric                                                         & Percentage \\ \hline
			Has invariants?                                                            & 92\,\%     \\
			Has discrete variables?                                                            & 24\,\%     \\
			Has multi-rate clocks                               & 17\,\%     \\
			L/U subclass                                                               & 19\,\%     \\
			Has silent actions?                                                        & 67\,\%     \\
			Strongly deterministic?                                                 & 78\,\%     \\ \hline
		\end{tabular}
	\end{subtable}
	\end{table}

\ea{At the end: give statistics on percentage of benchmarks with each syntactic extension (\eg{} invariants, stopwatches, multi-rate clocks), if-then-else…} \dm{I add \cref{table:statistics}. Maybe need to reformat it and add a description paragraph. I could add some metric statistics if they are output by the res file (here, only and all the metrics displayed in the webpage are kept)}

In \cref{table:statistics}, we present some statistics on our benchmarks.
Because of the presence of 3 benchmarks and 25 models (all in the ``monitoring'' category) with a very large number of locations (up to several dozens of thousands), only giving the average of some metrics is irrelevant.
To this end, we also provide the \emph{median} values.
Moreover, the average and the median of the number of discrete variables are computed only on the benchmarks which contains at least 1 such variable; they represent 24\% of our models.

\subsection{Properties}\label{subsection:properties}

Properties follow the \imitator{} syntax.
In the 1.0 version, they mainly consisted of reachability/safety properties; in addition, the properties were not explicitly provided, since \imitator{} 2.x did not specify properties (they were provided using options in the executed command).
In the new version of our library, we added several \emph{liveness} (cycle synthesis) properties, \ie{} for which one aims at synthesizing parameter valuations featuring at least one infinite (accepting) run~\cite{NPP18,AAPP21};
	in addition, we added properties such as deadlock-freeness synthesis (``exhibit parameter valuations for which the model is deadlock-free'')~\cite{Andre16}, optimal-parameter or minimal-time reachability~\cite{ABPV19}, and some ``pattern''-based properties~\cite{Andre13ICECCS} that eventually reduce to reachability checking~\cite{ABBL03}.

\LongVersion{
More in details, we provide and study ten\todo{} properties:
\begin{itemize}
	\item Safety (AGnot) to obtain the set of valuations for which the target location is unreachable.
	\item Cycle (resp.\ CycleThrough) to synthesize a parameter constraint such that, for any parameter valuation in that constraint, the system contains at least one infinite run \cite{NPP18,AAPP21} (resp.\ passing infinitely often through the specified location).
	\item DeadlockFree which synthesizes a parameter constraint such that, for any parameter valuation in that constraint, the system is deadlock-free \cite{Andre16}.
	\item Reachability (EF) to compute the set of parameter valuations for which some location is reachable.
	\item EFpmin (resp.\ EFpmax) to synthesize the minimum (resp.\ maximum) valuation for a given parameter for which a given location is reachable \cite{ABPV19}.
	\item EFtmin to synthesize the parameter valuation for which a given location is reachable in minimal time \cite{ABPV19}.
	\item Inverse method (IM) to compute a parameter constraint such that, for any parameter valuation in that constraint, the set of traces is the same as for the reference valuation \cite{ALM20}.\ea{add for future version: \cite{ACEF09}}
	\item Pattern.
\end{itemize}
}

\subsection{Unsolvable Benchmarks}\label{subsection:unsolvable}
 \begin{figure}[tb]
 	\centering
 	\scriptsize
 	\begin{subfigure}[b]{0.4\textwidth}

			\begin{tikzpicture}[pta, scale=1]

			\node[location, initial] at (0,0) (l1) {$\locinit$};
			\node [invariant, below] at (l1.south) {$\clockx \leq 1$};

			\node[location, accepting] at (3, 0) (lGoal) {$\locfinal$};

			\path (l1) edge[loop above] node[left, yshift=-.6em, align=center]{$ 	\clockx = 1$\\
				$\clockx \assign 0$\\
				} (l1);

			\path (l1) edge node[above,align=center] {$ \clockx = 0 \land  \clocky = \styleparam{p}$}
			(lGoal);
			\end{tikzpicture}

 		\caption{$\styleparam{p}\in\setN$}
 		\label{fig:unsolvable-N}
 	\end{subfigure}
 	~
 	\begin{subfigure}[b]{0.4\textwidth}
		\begin{tikzpicture}[pta, scale=1]

		\node[location, initial] at (0,0) (l1) {$\locinit$};
		\node [invariant,below] at (l1.south) {$ \clockx \leq 1$};

		\node[location, accepting] at (3,0) (lGoal) {$\locfinal$};

		\path (l1) edge[loop above] node[left, yshift=-.4em, align=center]{
			$\clockx = \styleparam{p}$ \\
			 $\clockx \assign 0$} (l1);

		\path (l1) edge node[align=center]{$ \clockx = 0 \land  \clocky = 1$
			} (lGoal);

		\end{tikzpicture}
 		\caption{$\styleparam{p}\in \left\{\frac1n, n\in\setN^*\right\}$}
 		\label{fig:unsolvable-InvN}
 	\end{subfigure}
	~
 	\begin{subfigure}[b]{0.4\textwidth}
		\begin{tikzpicture}[pta, scale=1]

		\node[location, initial] at (0,0) (l1) {$\locinit$};

		\node[location, accepting] at (3, 0) (lGoal) {$\locfinal$};

		\path (l1) edge[loop above] node[left, yshift=-.5em, align=center]{
			$ \clockx = \styleparam{p}$ \\
			 $\clockx \assign 0$
			} (l1);

		\path (l1) edge node[align=center]{$ \clockx = \styleparam{q} \land \clocky = \styleparam{r}$} (lGoal);

		\end{tikzpicture}
 		\caption{$\styleparam{r} \in \left\{n\times\styleparam{p}+\styleparam{q}, n\in\setN\right\}$}
 		\label{fig:unsolvable-pNplusq}
 	\end{subfigure}
 	~
 	\begin{subfigure}[b]{0.4\textwidth}
 		\begin{tikzpicture}[pta, scale=1]

 		\node[location, initial] at (0,0) (l1) {$\locinit$};
 		\node [invariant,below] at (l1.south) {\begin{tabular}{@{} c @{\ } c@{} }& $ \clockx \leq 1$\\\end{tabular}};

 		\node[location, accepting] at (3, 0) (lGoal) {$\locfinal$};

 		\path (l1) edge[loop above] node[left, yshift=-.6em, align=center]{\begin{tabular}{@{} c @{\ } c@{} }
 			& $ \clockx = 1$\\
 			& $\clockx \assign 0$\\
 			\end{tabular}} (l1);

 		\path (l1) edge node{\begin{tabular}{@{} c @{\ } c@{} }
 			& $ \clocky \geq \styleparam{p} \land \clockx = 0$\\
 			\end{tabular}} (lGoal);
 		\end{tikzpicture}
 		\caption{$\styleparam{p}\in \setRplus$}
 		\label{fig:unsolvable-Rplus}
 	\end{subfigure}
 	\caption{Examples of unsolvable benchmarks}\label{fig:unsolvables}
 \end{figure}
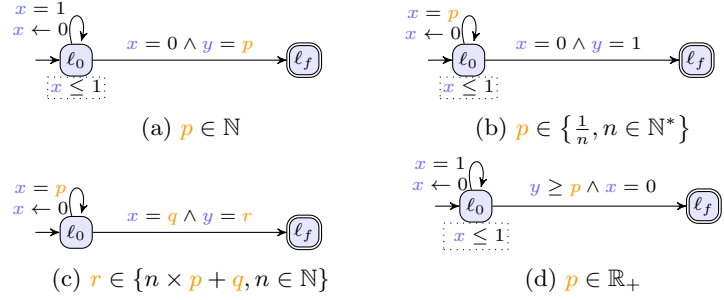

A novelty of our library is to provide a set of toy unsolvable benchmarks.
They have been chosen for being beyond the limits of the state-of-the-art techniques.
Four of them are illustrated in \cref{fig:unsolvables}.
For example, in \cref{fig:unsolvable-N}, the reachability of $\locfinal$ is achievable only if $\styleparam{p} \in \setN$; but no verification tool---to the best of our knowledge---terminates this computation.
Moreover, the final location of the PTA presented in \cref{fig:unsolvable-Rplus} is reachable for all $\styleparam{p}\geq0$, which is a convex constraint, but this solution remains not computable.
\subsection{Expected Performances}\label{subsection:performances}
Another novelty of the 2.0~version is to provide users with all the expected results, as generated by \imitator{}.
For all properties, we provide either a computed result, or (for the ``unsolvable'' benchmarks), a human-solved theoretical result.

We also give an approximate execution time, and the number of (symbolic) states explored.
These metrics are not strict, as they may depend on the target model checker and the target hardware/OS, but this provides the user an idea of the complexity of our models.

\LongVersion{
If \imitator{} is able to compute the result of a property application, we provide its output file and extract some relevant features to have a general idea of its performances: the total computation time, the number of states and the number of computed states. The executions were made on an Intel Xeon Gold 5220 CPU @ 2.20\,GHz with 96\,GiB running Linux Ubuntu 20.

If \imitator{} is not able to compute it---recall that the concerned benchmarks are tagged with \enquote{Unsolvable}---, we provide a similar result file, whose extension is \texttt{expres}. In this case, the metrics as presented as \enquote{Not executed (Unsolvable)} (\enquote{NE (Uns.)}).
}

\begin{table}[tb]
	\footnotesize
	\caption{Statistics on executions (over 157 properties)}\label{table:statistics-executions}
	\centering
\begin{tabular}{| l| c| c|}
	\hline
	\rowHeader{}Metric        & Average & Median \\ \hline
	Total computation time    & 245.8   & 2.819  \\
	Number of states          & 20817.8 & 580  \\
	Number of computed states & 34571.7 & 1089 \\ \hline
\end{tabular}
\end{table}

In \cref{table:statistics-executions}, we present the statistics over 157 imitator executions.
Note that the \emph{unsolvable} executions \LongVersion{(which are computed with a timeout) }are not included in this table.

\section{Perspectives}\label{section:perspectives}

\LongVersion{
To allow a version of our library in both the \imitator{} format and the JANI format, we implemented within \imitator{} the translation of an \imitator{} model to its JANI specification.
In order for \imitator{} to also be fed with other existing models specified in JANI, we would like to implement the reverse translation (from JANI to \imitator{}), using JANI tools or within \imitator{}.
\ea{not sure whether we should add this; this concerns more \imitator{} itself than the library?}
}

Ultimately, we hope our library can serve as a basis for a \emph{parametric} timed model checking competition, a concept yet missing in the model checking community.

Opening the library to volunteer contributions is also on our agenda.

 \section*{Acknowledgements}
Experiments presented in this paper were carried out using the Grid'5000 testbed, supported by a scientific interest group hosted by Inria and including CNRS, RENATER and several Universities as well as other organizations (see \href{https://www.grid5000.fr}{https://www.grid5000.fr}).

\ifdefined\VersionWithComments
	\newcommand{\CCIS}{Communications in Computer and Information Science}
	\newcommand{\ENTCS}{Electronic Notes in Theoretical Computer Science}
	\newcommand{\FAC}{Formal Aspects of Computing}
	\newcommand{\FI}{Fundamenta Informaticae}
	\newcommand{\FMSD}{Formal Methods in System Design}
	\newcommand{\IJFCS}{International Journal of Foundations of Computer Science}
	\newcommand{\IJSSE}{International Journal of Secure Software Engineering}
	\newcommand{\IPL}{Information Processing Letters}
	\newcommand{\JAIR}{Journal of Artificial Intelligence Research}
	\newcommand{\JLAP}{Journal of Logic and Algebraic Programming}
	\newcommand{\JLAMP}{Journal of Logical and Algebraic Methods in Programming} %
	\newcommand{\JLC}{Journal of Logic and Computation}
	\newcommand{\LMCS}{Logical Methods in Computer Science}
	\newcommand{\LNCS}{Lecture Notes in Computer Science}
	\newcommand{\RESS}{Reliability Engineering \& System Safety}
	\newcommand{\STTT}{International Journal on Software Tools for Technology Transfer}
	\newcommand{\TCS}{Theoretical Computer Science}
	\newcommand{\ToPNoC}{Transactions on Petri Nets and Other Models of Concurrency}
	\newcommand{\TSE}{{IEEE} Transactions on Software Engineering}
\else
	\newcommand{\CCIS}{CCIS}
	\newcommand{\ENTCS}{ENTCS}
	\newcommand{\FAC}{FAC}
	\newcommand{\FI}{FI}
	\newcommand{\FMSD}{FMSD}
	\newcommand{\IJFCS}{IJFCS}
	\newcommand{\IJSSE}{IJSSE}
	\newcommand{\IPL}{IPL}
	\newcommand{\JAIR}{JAIR}
	\newcommand{\JLAP}{JLAP}
	\newcommand{\JLAMP}{JLAMP}
	\newcommand{\JLC}{JLC}
	\newcommand{\LMCS}{LMCS}
	\newcommand{\LNCS}{LNCS}
	\newcommand{\RESS}{RESS}
	\newcommand{\STTT}{STTT}
	\newcommand{\TCS}{TCS}
	\newcommand{\ToPNoC}{ToPNoC}
	\newcommand{\TSE}{TSE}
\fi

\ifdefined\VersionAuthor
	\renewcommand*{\bibfont}{\small}
	\printbibliography[title={References}]
\else
	\bibliographystyle{splncs04} %
	\bibliography{PTA}
\fi

\LongVersion{
\appendix
\section{An IPTA illustrative example}\label{appendix:example}

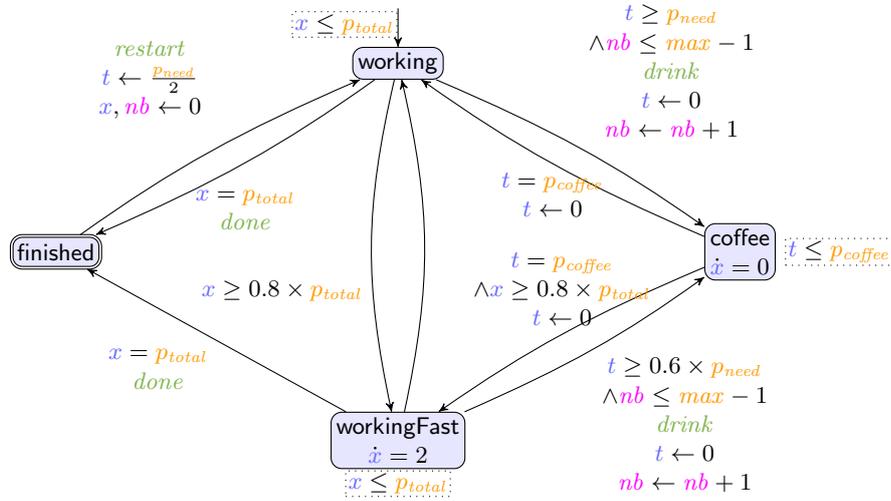
\begin{figure}[htb]

	\centering
	 \footnotesize

	 \scalebox{1}{
	\begin{tikzpicture}[pta, scale=1, xscale=2.25, yscale=1.25, bend angle=7]

		\node[location, initial, initial where=above] at (0, 0) (working) {\styleloc{working}};

		\node[location, accepting] at (-2, -2) (finished) {\styleloc{finished}};

		\node[location] at (0, -4) (workingFast) [align=center] {\styleloc{workingFast}\\$\dot{\clockx} = 2$};

		\node[location] at (2, -2) (coffee) [align=center] {\styleloc{coffee} \\ $\dot{\clockx} = 0$};

		\node[invariant, above=of working, yshift=-25, xshift=-20] {$\clockx \leq \paramTotal$};
		\node[invariant, below=of workingFast, yshift=28] {$\clockx \leq \paramTotal$};
		\node[invariant, right=of coffee, xshift=-25] {$\clockt \leq \paramCoffee$};

		\path (working) edge[bend right] node[below left,align=center, yshift=-10]{$\clockx \geq 0.8 \times \paramTotal$} (workingFast);
		\path (workingFast) edge[bend right] node[align=center]{} (working);

		\path (working) edge[bend left] node[below, align=center, yshift=-7]{$\clockx = \paramTotal$\\\actDone{}} (finished);
		\path (finished) edge[bend left] node[above left, align=center]{%
			\actRestart{}
			\\
			$\clockt \assign \frac{\paramNeed}{2}$
			\\
			$\clockx , \nb \assign 0$
			\\} (working);

		\path (workingFast) edge[] node[align=center]{$\clockx = \paramTotal$\\\actDone{}} (finished);

		\path (working) edge[bend left] node[above right, align=center]{%
			$\clockt \geq \paramNeed$\\$\land \nb \leq \MAXCOFFEE - 1$
			\\
			\actDrink{}
			\\
			$\clockt \assign 0$
			\\
			$\nb \assign \nb+1$} (coffee);
		\path (coffee) edge[bend left] node[below, align=center]{$\clockt = \paramCoffee$\\$\clockt \assign 0$} (working);

		\path (workingFast) edge[bend right] node[below right, align=center]{%
			$\clockt \geq 0.6 \times \paramNeed$\\$\land \nb \leq \MAXCOFFEE - 1$
			\\
			\actDrink{}
			\\
			$\clockt \assign 0$
			\\
			$\nb \assign \nb+1$} (coffee);
		\path (coffee) edge[bend right] node[above, align=center]{$\clockt = \paramCoffee $\\$\land \clockx \geq 0.8 \times \paramTotal$\\$\clockt \assign 0$} (workingFast);

	\end{tikzpicture}
	}
	\caption{An \enquote{\imitator{} PTA} example: Writing papers and drinking coffee}
	\label{figure:example-researcher}

\end{figure}
\begin{example}\label{example:researcher:full}
Consider the IPTA (\enquote{\imitator{} PTA}) in \cref{figure:example-researcher}\footnote{%
	This case study \benchmark{researcher} is part of the version~2 of our library, in categories ``academic'', ``toy'' and ``teaching''.
}, modeling a researcher writing papers~\cite{Andre21}.
The model features two clocks $\clockt$ (measuring the time when needing a coffee) and $\clockx$ (measuring the amount of work done on a given paper), both initially~0.
Their rate is always~1, unless otherwise specified (\eg{} in \styleloc{workingFast}).
Initially, the researcher is working (location \styleloc{working}) on a paper, requiring an amount of work $\paramTotal$.
When the paper is completed (guard $\clockx = \paramTotal$), the IPTA moves to location $\styleloc{finished}$.
From there, at any time, the researcher can start working on a new paper (transition back to \LongVersion{location }\styleloc{working}, updating $\clockx$ and~$\clockt$).

Alternatively, after at least a certain time (guard $\clockt \geq \paramNeed$), the researcher may need a coffee; this action can only be taken until a maximum number of coffees have been drunk for this paper ($\nb \leq \MAXCOFFEE - 1$), where $\nb$ is a discrete global variable recording the number of coffees drunk while working on the current paper.
When drinking a coffee (location \styleloc{coffee}), the work is obviously not progressing ($\dot{\clockx } = 0$).
Drinking a coffee takes exactly $\paramCoffee$ time units (guard $\clockt = \paramCoffee$ back to location \styleloc{working}).
Observe that, from the second paper onward (transition labeled with \actRestart{}), the researcher is already half-way of her/his need for a coffee (parametric update $\clockt \assign 0.5 \times \paramNeed$ \cite{ALR19FORTE}).

Also, whenever 80\,\% of the work is done (guard $\clockx \geq 0.8 \times \paramTotal$), the researcher may work twice as fast (location \styleloc{workingFast}, with a rate~2 for clock~$\clockx$).
In that case, (s)he needs a coffee faster too ($0.6 \times \paramNeed$).

All three durations $\paramCoffee$, $\paramNeed$ and $\paramTotal$ are timing parameters.
We fix their parameter domains as follows: $\paramCoffee,\paramTotal \in [0, \infty)$ and $\paramNeed \in [1, \infty)$.
The maximum number of coffees $\MAXCOFFEE \in [0, \infty)$ is also a parameter; observe that it is (only) compared to the discrete variable~$\nb$, which is allowed by the liberal syntax of \imitator{}.

\end{example}

}

\end{document}